\documentclass[aps,twocolumn,amssymb,superscriptaddress]{revtex4-1} 
\usepackage[final]{changes}
\usepackage[colorlinks=true,citecolor=blue,linkcolor=blue,urlcolor=blue]{hyperref}
\usepackage{amsmath}
\usepackage{amssymb}
\usepackage{bbm}
\usepackage{xcolor}
\usepackage{graphicx}
\usepackage{epstopdf}
\usepackage{dcolumn}
\usepackage{bm}
\usepackage{hyperref}
\graphicspath{{Latexpicture/}}

\begin{document}
\title{Ultrafast adiabatic passages in ultrastrongly coupled light-matter systems}

\author{Xun Gao}
\affiliation{Hunan Key Laboratory for Micro-Nano Energy Materials and Devices\\ and School of
Physics and Optoelectronics, Xiangtan University, Hunan 411105, China}

\author{Liwei Duan}
\affiliation{Department of Physics, Zhejiang Normal University, Jinhua 321004, China}

\author{Pinghua Tang}
\affiliation{Hunan Key Laboratory for Micro-Nano Energy Materials and Devices\\ and School of
Physics and Optoelectronics, Xiangtan University, Hunan 411105, China}

\author{Junlong Tian}
\affiliation{Department of Electronic Science, College of Big Data and Information Engineering, Guizhou University, Guiyang 550025, China}

\author{Zhongzhou Ren}
\affiliation{School of Physics Science and Engineering, Tongji University, Shanghai 200092, China}

\author{Enrique Solano}
\email{enr.solano@gmail.com}
\affiliation{Kipu Quantum, Greifswalderstrasse 226, 10405 Berlin, Germany}

\author{Jie Peng}
\email{jpeng@xtu.edu.cn}
\affiliation{Hunan Key Laboratory for Micro-Nano Energy Materials and Devices\\ and School of
Physics and Optoelectronics, Xiangtan University, Hunan 411105, China}
\begin{abstract}
We have obtained the solutions of the multimode quantum Rabi model when all modes have identical frequencies $\omega$, including dark states $|\phi_K\rangle$ with at least $K$ $(K=1,2,3,\ldots)$ photons. Extended to the multiqubit case, they lie close to another dark state $\vert \psi\rangle$ with at most one photon in the spectrum. Taking advantages of such solutions, we find a linear and symmetry-protected adiabatic passage through $\vert \psi\rangle$ to fast generate arbitrary single-photon $M$-mode $W$ states $\vert W_M\rangle$ with exactly the same speed. The effective minimum energy gap during the adiabatic evolution is further enlarged to $0.63\omega$ when Stark shifts are included, such that arbitrary $\vert W_M\rangle$ can be ultrafast generated in $1.55\times 2\pi\omega^{-1}$ with fidelity $99\%$, indepedent of $M$. This work reveals the existence of linear ultrafast adiabatic passages in light-matter systems.  
\end{abstract}
\maketitle
\emph{Introduction.--}
The main advantage of adiabatic evolution in performing population transfers is its robustness against small variations of experimental conditions \cite{sti}, compared to resonant pulses. The price to pay is longer interaction time, negatively related to the minimum energy gap $\Delta E_{\min}$ during the adiabatic evolution \cite{xxy}. $\Delta E_{\min}$ is proportional to the coupling strength $g$ between light and matter in typical adiabatic passages that invert the population of the qubit, e. g., the Landau-Zener scheme \cite{lz}, where the detunning changes linearly and $g$ is a constant \cite{chen xi}. Since $g$ is much smaller than the photon frequency $\omega$ to satisfy the rotating-wave approximation, such simple linear adiabatic path results in slow adiabatic speed, and many approaches are proposed to overcome this disadvantage, like path optimization \cite{xxy,martins1,martins2,pancho}, and shortcut to adiabaticity \cite{chen xi,chenxi1,short}. However, they need delicate control of pulse shaping, or additional transitions.

A way to linear ultrafast adiabatic passage is simply to increase $g$. However, counter-rotating terms will be involved when we enter the ultrastrong coupling regime $g\gtrsim 0.1\omega$ \cite{kike1,kike2,nori3}. The system described by the quantum Rabi model \cite{otp,sqi,jc} (QRM) consists of infinite photons, which are entangled with the qubit, making it impossible to flip the qubit. 
Its solution \cite{iot,chen,hzhong}, dynamics \cite{jorge, wolf} and other generalizations are extensively studied  \cite{irish,hwang,liu,Iorsh,jie,flecti,ying,zhang1,zhang2,luoh,lv,adrian,ultrafast}.
Recently, a special dark-state solution with at most one photon $\vert \psi_{2+}\rangle$ was found in the two-qubit multimode quantum Rabi model (MQRM) \cite{oso}. It can be used to fast generate a two-qubit Bell state and a single-photon $M$-mode $W$ state through linear adiabatic path, which are both important resources in quantum information \cite{agw,zheng,67 (2018),042102 (2002),014302 (2006),054302 (2002),812(2015),yanxia1,yanxia2,prl 2020,circuit1,stoj,spin,cole,chiral1}.
However, the ultrafast state-generation is not realized and the mechanism behind the fast adiabatic speed is still unclear.
Counterintuitively, $\Delta E_{\min}$ is vanishing small during some part of the adiabatic evolution.
The full solutions to the MQRM will reveal its possible symmetry protecting the adiabatic evolution along the peculiar one-photon solution, and may show a way to further enlarge the effective $\Delta E_{\min}$, leading to ultrafast linear adiabatic passages.

Here we have obtained the full solutions to the MQRM when all modes have identical frequencies $\omega$, including dark states $\vert \phi_K\rangle$ with at least $K ~(K=1,2,3,\ldots)$ photons, related to conserved bosonic number operators. Generalized to the multiqubit case, $\vert\phi_K\rangle$ lies close to $\vert \psi_{2+}\rangle$ with vanishing small gap in some part of the spectrum, but belongs to different subspaces, so that it will not affect the adiabatic evolution along $\vert \psi_{2+}\rangle$.
More interestingly, the spectrum of the $M$-mode QRMs are exactly the same for different $M$ if $\sum_{i=1}^M g_{i}^2$ is fixed, except for energy levels $\vert \phi_K\rangle$, where $g_i$ is the coupling strength between the qubit and the $i$-th mode. Accordingly, we can tune  $g_i/g_{i'}$ while keep $\sum_{i=1}^M g_{i}^2$ fixed to generate arbitrary $\vert W_M\rangle=(1/{\cal{N}})\sum_{i=1}^M g_i\vert 0_10_2\ldots1_i0_{i+1}\ldots0_M\rangle$ from vacuum, through adiabatic evolution along $\vert \psi_{2+}\rangle$ with exactly the same speed, independent of $M$, because the effective energy gap and dynamics will be the same.

However, the effective $\Delta E_{\min}$ is still not large enough for ultrafast state-generation. Therefore, we study the multiqubit MQRM with the addition of Stark shifts, and find another kind of dark-state solutions with at most one photon and constant eigenenergy $E=\omega$ in the whole coupling regime, for arbitrary qubit and mode numbers, where $\vert \psi_{2s+}\rangle$ can be used to ultrafast generate $\vert W_M\rangle$ through linear adiabatic path. Although other energy levels $\vert E_m\rangle$ consists of infinite photons, only the one-photon part will affect the adiabatic speed. Moreover, $\langle \phi_K|\dot{H}|\psi_{2s+}\rangle=0$, and $\langle E_m=\omega+\delta|\dot{H}|\psi_{2s+}\rangle\propto|\delta|f(|\delta|)$, which decreases as $\vert E_m\rangle$ goes closer to $\vert \psi_{2s+}\rangle$. So that the effective $\Delta E_{\min}$  reaches $0.63\omega$ and  arbitrary $\vert W_M\rangle$ can be ultrafast generated with the same fidelity $99\%$ in $1.55\times 2\pi\omega^{-1}$ by fixing $\sum_{i=1}^M g_{i}^2$.

\emph{Solutions to the MQRM when all modes have identical frequencies.}--
First, we study the MQRM
\begin{eqnarray}\label{h2m}
H=\Delta\sigma_{z}+\sum_{i=1}^{M}\omega_i a_i^\dagger a_i +g_{i}\sigma_{x}(a_i^\dagger+a_i),
\end{eqnarray} 
where $a_i^\dagger (a_i)$ are the $i$-th photon mode creation (annihilation) operators with frequency $\omega_i$ . The qubit is described by Pauli matrices $\sigma_x$ and $\sigma_z$ with energy level splitting $2\Delta$. 

This model is solvable when $\omega_i=\omega$, because we can apply Bogoliubov transformations
\begin{eqnarray}\label{bogo}
b_1&=&\frac{\sum_{i=1}^M g_i a_i}{\sqrt{\sum_{i=1}^M g_i^2}},\nonumber\\
b_j&=&\frac{\sum_{i=1}^{j-1}g_ig_ja_i-\sum_{i=1}^{j-1} g_i^2 a_j}{\sqrt{\sum_{i=1}^j g_i^2\sum_{i=1}^{j-1} g_i^2}},~~~j=2,3,\ldots,M,
\end{eqnarray}
to transform $H$ into
\begin{eqnarray}\label{h2m}
H'&=&H'_R+\sum_{j=2}^{M}\omega b_j^\dagger b_j,\nonumber\\
H'_R&=& \Delta\sigma_{z}+\omega b_1^\dagger b_1+(\sum_{i=1}^M g_i^2)^{\frac{1}{2}}\sigma_{x}(b_1^\dagger+b_1),
\end{eqnarray} 
which is a combination of a single-qubit QRM $H'_R$ and $M-1$ free bosonic modes \cite{trans}, where the former has been solved \cite{iot,chen}.  Note that the transformations Eq. \eqref{bogo} reduces to the unitary transformation Chilingaryan and Rodr\'{i}guez-Lara used for the two-mode case \cite{cr}.  $H'$ will bring interesting and useful results. First, with the solution to the one-mode Rabi model at hand \cite{iot,chen,hzhong}, we just need to replace $g$ with $(\sum_{i=1}^M g_i^2)^{\frac{1}{2}}$ and  $(a^\dag)^n\vert 0\rangle$ with $(b_1^\dag)^n\vert 0\rangle$ to obtain the solution to $H'_R$, $\vert \psi_R^\prime\rangle$. E. g., $\vert 1\rangle$ will become $\vert W_M\rangle=(\sum_{i=1}^M g_i^2)^{-\frac{1}{2}}\sum_{i=1}^M g_i\vert 0_10_2\ldots1_i0_{i+1}\ldots0_M\rangle$.
Second, the solution to $H'$ takes the form of $\vert \psi_R^\prime\prod_{j=2}^M n_{b_j}\rangle$, where $\vert n_{b_j}\rangle$ is the eigenstate of $ b_j^\dagger b_j$. $\vert \psi_R^\prime\rangle$ consists of all Fock states. However, once a $n_{b_{j}}\neq 0$, the solution $\vert\phi_K\rangle$ will have at least $K=\sum_{j>1} n_{b_{j}}$ photons. Although $H$ connects all the Fock states, these dark states satisfy $\langle K-1|H|\phi_K\rangle=0$, because the coherent superposition of basis with $K$ photons. Taking the two-mode case for example, the zero-photon part $\vert \uparrow 00\rangle$ vanishes when a solution with at least one photon $\vert \downarrow \rangle(g_2\vert 10\rangle-g_1\vert 01\rangle)+\ldots$ is applied by $H$. And $g_2\vert 10\rangle-g_1\vert 01\rangle$ just corresponds to $b_2^\dag\vert 00\rangle$, so it is easy to obtain $\vert \phi_K\rangle$ using $\prod_j (b_j^\dag)^ {n_{b_j}}\vert 0\ldots0\rangle$ with $\sum n_{b_j}=K$. Third, the spectrum of the MQRMs are just the same for different $M$ when $(\sum_{i=1}^M g_i^2)^{\frac{1}{2}}$ is fixed, except for the addition of $\vert\phi_K\rangle$, whose energy is lifted by $K\omega$. Last, there are $M-1$ symmetries generated by $n_{b_j}$ beside $H$ and parity
$\exp(i\pi\sum a_i^\dag a_i )\sigma_z$, so there are level crossings between energy levels within the same parity subspace.

\emph{Extended to the multiqubit case for fast generation of arbitrary $W$ states with exactly the same speed.--}We generalize $H$ to the multiqubit MQRM
\begin{equation}\label{mode}
H_{pq}=\sum_{i=1}^M\omega_i a_i^\dagger a_i+\sum_{i=1}^M\sum_{j=1}^N g_{ij} \sigma_{jx}(a_i+a^\dagger_i)+\sum_{j=1}^N \Delta_j \sigma_{jz},
\end{equation}
to find similar properties, where $g_{ij}$ is the coupling strength between the $i$-th mode and $j$-th qubit. $2\Delta_j$ is the energy splitting of the $j$-th qubit. When $\omega_i=\omega$ and $\frac{g_{ij}}{g_{i'j}}$ is independent of $j$, $H_{pq}$ can be transformed into
\begin{eqnarray}\label{hpqp}
H'_{pq}&=&\omega b_1^\dag b_1+\sum_{j}(\sum_{i}g_{ij}^2)^{\frac{1}{2}}\sigma_{jx}(b_1^\dag+b_1)+\sum_{j=1}^N\Delta_j\sigma_{jz}\nonumber\\&&+\omega\sum_{i=2}^M b_i^\dag b_i,
\end{eqnarray}  
after the same transformation Eq. \eqref{bogo}. So the solution to $H'_{pq}$ still has the form of $\vert \psi^\prime_{qR}\prod_{i=2}^M n_{b_i}\rangle$, where $\vert \psi^\prime_{qR}\rangle$ is the solution of the single-mode multiqubit QRM. Using this relation, we can recover the one-photon solutions to the two-qubit MQRM \cite{oso}
\begin{equation}\label{psi2+m}
\vert \psi_{2+}\rangle=\frac{1}{\cal{N}}[(\Delta_1-\Delta_2)\vert 0_M\uparrow\uparrow\rangle+(\sum_{i}g_{i}^2)^{\frac{1}{2}}\vert W_M(\downarrow\uparrow-\uparrow\downarrow)\rangle],
\end{equation}
which exists when $\Delta_1+\Delta_2=\omega$, $g_{i1}=g_{i2}=g_i$, from the one-photon solution to the two-qubit QRM \cite{sot,jcg}
\begin{equation}\label{psid}
\vert \psi_d\rangle=\frac{1}{\cal{N}}[(\Delta_1-\Delta_2)\vert 0\uparrow\uparrow \rangle+g\vert 1(\downarrow\uparrow-\uparrow\downarrow)\rangle],
\end{equation}
by replacing $g$ with $(\sum_{i}g_{i}^2)^{\frac{1}{2}}$, $\vert 1\rangle$ with $\vert W_M\rangle$, and choosing $\sum_{i=2}^M n_{b_i}=0$. It exists when $g=g_{11}=g_{12}$, $\Delta_1+\Delta_2=\omega$. Both solutions have constant energy $E=\omega$ in the whole coupling regime.

\begin{figure}[tb]
\centering
\includegraphics[scale=0.39]{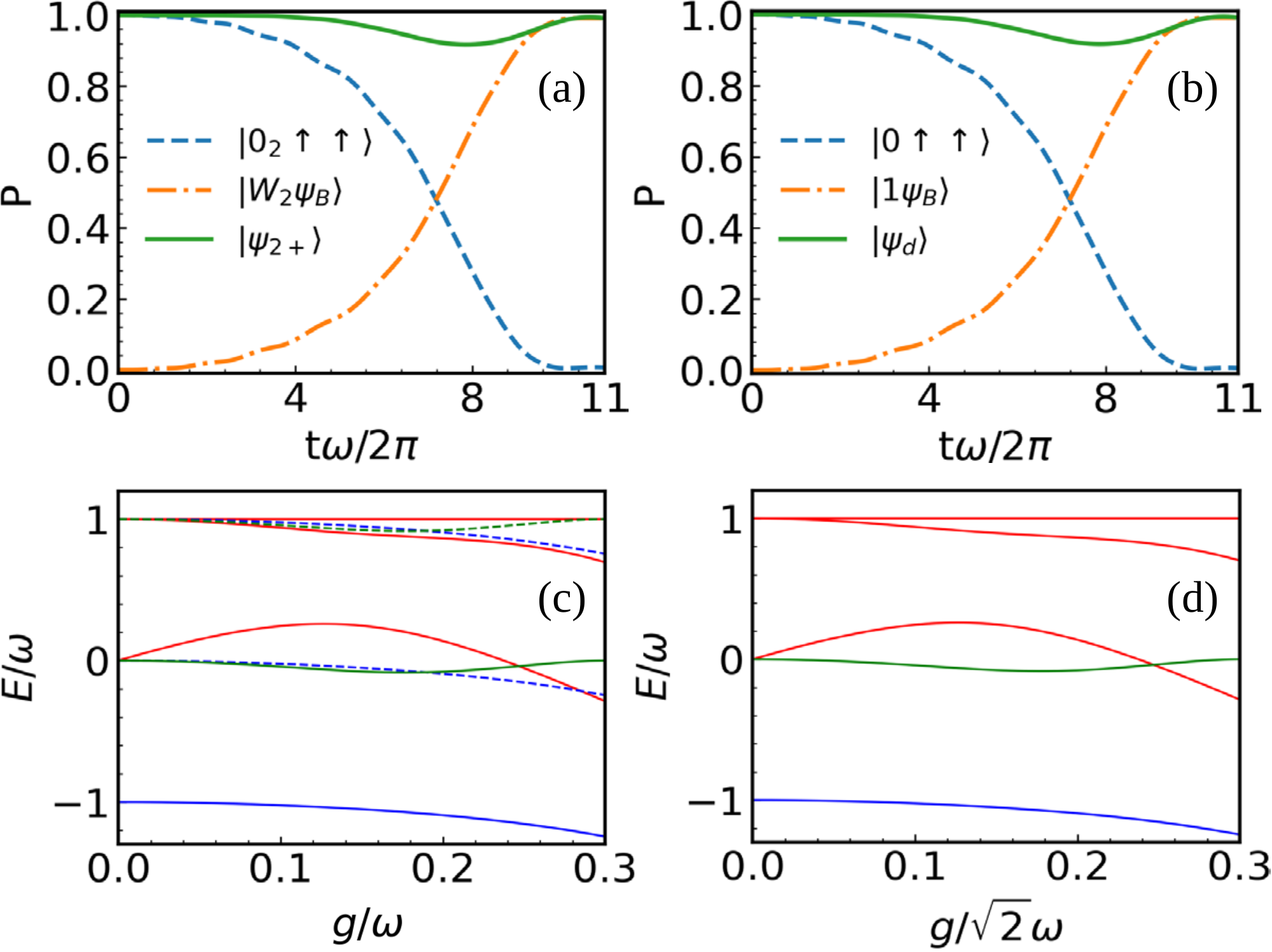}
\renewcommand\figurename{\textbf{FIG.}}
\caption[1]{(a) Generating $\vert W_2\psi_B\rangle$ from $\vert 0_2\uparrow\uparrow\rangle$ through adiabatic evolution along $\vert \psi_{2+}\rangle$, where $\vert\psi_B\rangle=\vert \downarrow\uparrow-\uparrow\downarrow\rangle/\sqrt{2}$. $\Delta_1(t=0)=\omega$, $\Delta_2(t=0)=0$, $g_1(t=0)=g_2(t=0)=0$, $\dot{\Delta}_2=-\dot{\Delta}_1=0.5/(11\times 2\pi)\omega^{2}$, $\dot{g}_1=\dot{g}_2=0.3/(11\times 2\pi)\omega^{2}$. (b) Generating $\vert 1\psi_B\rangle$ from $\vert 0\uparrow\uparrow\rangle$ through adiabatic evolution along $\vert \psi_{d}\rangle$. The parameters are the same as in (a) except $g(t)=\sqrt{2}g_1=\sqrt{2}g_2$. (c) Instantaneous spectrum of the two-qubit two-mode QRM during the adiabatic process in (a). (d) Instantaneous spectrum of the two-qubit QRM during the adiabatic process in (b).}
\label{figa}
\end{figure}

$\vert \psi_{2+}\rangle$ and $\vert \psi_d\rangle$ can be used to generate arbitrary $\vert W_M\rangle$ and $\vert 1\rangle$ respective, with exactly the same speed through adiabatic evolution, independent of $M$, as shown in Figs. \ref{figa} (a) and (b). The reason is as follows:
To generate  $\vert W_M\rangle=\frac{1}{\cal{N}}\sum_{i=1}^M g_i\vert 0_10_2\ldots1_i0_{i+1}\ldots0_M\rangle $ along $\vert \psi_{2+}\rangle$ (Eq. \eqref{psi2+m}) from $\vert 0_M\uparrow\uparrow\rangle$ adiabatically, we set $g_{ij}=g_{i}=0$, $\Delta_1+\Delta_2=\omega$, $\Delta_1\neq\Delta_2$ initially, and $\dot{g}_i/\dot{g}_{i'}=g_i/g_{i'}$, $\dot{\Delta}_1=-\dot{\Delta}_2$, so
\begin{eqnarray}\tiny
\dot{H}(t)&=&\dot{\Delta}_1(\sigma_{1z}-\sigma_{2z})+\sum_i \dot{g}_i(a_i^\dag+a_i)(\sigma_{1x}+\sigma_{2x})
\\
&=&\dot{\Delta}_1(\sigma_{1z}-\sigma_{2z})+(\sum_{i}g_{i}^2)^{\frac{1}{2}}\frac{\dot{g_i}}{g_i}(b_1+b_1^\dagger)(\sigma_{1x}+\sigma_{2x}).\nonumber
\end{eqnarray}
$\vert W_M\psi_B\rangle$ is generated when $\Delta_1=\Delta_2$, where $\vert\psi_B\rangle=\vert \downarrow\uparrow-\uparrow\downarrow\rangle/\sqrt{2}$. $H(t)$ acts only on the subspace of $\{\vert n_{b_1}\rangle\}$, and the dynamics of the $M$-mode multiqubit QRM is exactly the same in this subspace if $\sum_{i=1}^M g_{ij}^2$ is fixed, as can be seen in $H'_{pq}$ (Eq. \eqref{hpqp}), so $\vert W_M\psi_B\rangle$ can be generated with exactly the same speed, independent of $M$. This is testified by the numerical simulation shown in Figs. \ref{figa} (a) and (b), where
$\vert W_2\psi_B\rangle$ and  $\vert 1\psi_B\rangle$ are fast generated from   $\vert 0\uparrow\uparrow\rangle$ in $t=11\times2\pi\omega^{-1}$ with fidelity $99.17\%$. If  $\omega=2\pi\times 3$GHz here and hereafter, then $t=3.67$ ns.

The fast adiabatic speed
can be explained from the adiabatic theorem \cite{amin,aiu,bda}, which states that the system will evolve adiabatically along $|E_{n}\rangle$ if
\begin{equation}\label{adia1}
\left|\frac{\langle E_{m}(t)|\dot{H}|E_{n}(t)\rangle}{(E_m-E_n)^2}\right|\ll1 , ~~~
m\neq n , ~~~ t\in[0,T].
\end{equation} 
We show the instantaneous  spectra of the two-qubit two-mode QRM and two-qubit single mode QRM during the adiabatic evolution, in Figs. \ref{figa} (c) and (d) respectively. Since  $\sum_{i=1}^M g_{ij}^2$ is fixed, they are exactly the same except for dashed lines with $n_{b_2}\neq0$, according to $H'_{pq}$ (Eq. \eqref{hpqp}). They are just translations of the solid lines with the same color by $n_{b_2}\omega$, and has at least $n_{b_2}$ photons. Two dashed lines $\vert \phi_K\rangle$ are close to $\vert \psi_{2+}\rangle$, but will not affect the adiabatic speed because  $\dot{H}(t)|\psi_{2+}\rangle$ has $n_{b_2}=0$ and hence $\langle \phi_K|\dot{H}(t)|\psi_{2+}\rangle=0$. There is still one solid line  $\vert E_m\rangle$ close to $\vert \psi_{2+}\rangle$ with vanishing small gap when $g\approx0$.  Although $\vert E_m\rangle$ consists of infinite basis, only $\vert 1_{b_1}\uparrow\downarrow\rangle$ and $\vert 1_{b_1}\downarrow\uparrow\rangle$ will affect the adiabatic speed and $\langle E_m=\omega+\delta|\dot{H}|\psi_{2+}\rangle\propto f(\Delta_{1,2},g_{i})|\delta|/\sqrt{(\Delta_1-\Delta_2)^2+\delta^2}$ \cite{sl}, which decreases as $\vert E_m\rangle$ goes closer to $|\psi_{2+}\rangle$. So that we chose the largest $\Delta_1-\Delta_2=\omega$ when $g= 0$ to ensure the fast generation of $\vert W_M\rangle$. Accordingly, both qubits are far detuned from the resonator when we excite the qubit, so that the crosstalk can be neglected. Moreover, since the effective spectra and dynamics are the same when  $\sum_{i=1}^M g_{ij}^2$ is fixed, the adiabatic speed to generate different $\vert W_M\rangle$ are guaranteed to be the same.

\emph{Linear ultrafast adiabatic passages to generate arbitrary $W$ states}--
However, the ultrafast state-generation remains unrealized so for.
Some energy levels are too close to $\vert \psi_{2+}\rangle$, e.g., $\vert 2_{b_1}0_{b_2}
\downarrow\downarrow\rangle$ and $\vert 0_{b_1}0_{b_2}\uparrow\uparrow\rangle$ are degenerate at $g=0$ since $\Delta_1+\Delta_2=\omega$. Intuitively, if we add Stark shift terms $U_{ij}\sigma_{jz} a_i^\dagger a_i$ to the Hamiltonian, then they could be separated in the spectrum. Meanwhile, the Rabi-Stark model can be realized in cavity QED \cite{cso,orm} with all parameters adjusted freely and
independently \cite{cso,ago}, or in trapped ions \cite{sii}, and are extensively studied in \cite{amo,aes,qrm,est}. So we study the multiqubit multimode Rabi-Stark model 
\begin{eqnarray}\label{stark}
H_S&=&\sum_j\Delta_j\sigma_{jz}+\sum_i(\omega_i+\sum_{j}U_{ij}\sigma_{jz})a_i^\dagger a_i \nonumber\\ &&+\sum_{ij}g_{ij}(a_i^\dagger+a_i)\sigma_{jx},
\end{eqnarray} and find similar one-photon solutions with constant eigenenergy $E=\omega_i=\omega$ in the whole coupling regime for arbitrary qubit and mode numbers \cite{sl}. Here parity $\exp(i\pi\sum_i a_i^\dag a_i)\prod_j\sigma_{jz}$ is still conserved. For the two-qubit $M$-mode case, such solution reads
\begin{equation}\label{dark1m}
	|\psi_{2s+}\rangle =\frac{1}{{\cal N}}\left[|0_M\uparrow\uparrow\rangle+|W'_M(\downarrow\uparrow-\uparrow\downarrow)\rangle\right]
\end{equation}
for even parity, with the condition $\omega_i=\omega=\Delta_1+\Delta_2$ and $g_{i1}=g_{i2}=g_i$ for $i$ from $1$ to $M$, where $|W'_M\rangle=\frac{g_1}{\Delta_1-\Delta_2+U_{11}-U_{12}}|100\cdots0\rangle+\frac{g_2}{\Delta_1-\Delta_2+U_{21}-U_{22}}|010\cdots0\rangle+\cdots+\frac{g_M}{\Delta_1-\Delta_2+U_{M1}-U_{M2}}|000\cdots1\rangle$. $|\psi_{2s+}\rangle$ can be used to generate arbitrary $W$ states $|W'_M\rangle$ through adiabatic evolution.

\begin{figure}[tb]
\centering
\includegraphics[scale=0.39]{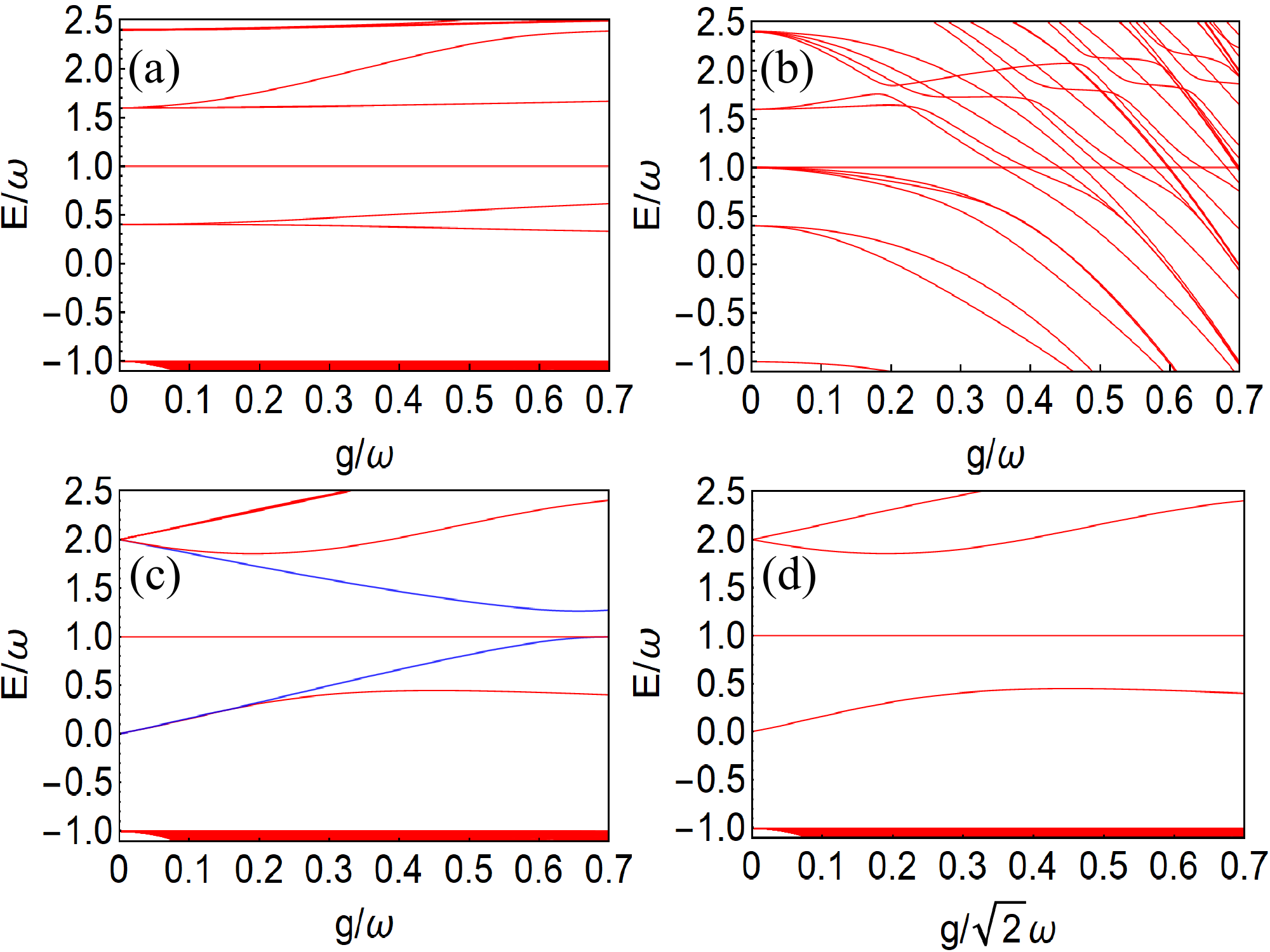}

\renewcommand\figurename{\textbf{FIG.}}
\caption[1]{(a) Even-parity spectrum of the two-qubit two-mode quantum Rabi-Stark model with $\Delta_1=0.8\omega,\Delta_2=0.2\omega,U_{11}=U_{12}=U_{21}=U_{22}=0.5\omega$ and $g_{11}=g_{12}=g_{21}=g_{22}=g$ for even parity. (b) Even-parity spectrum of the two-qubit two-mode QRM with the same parameters as (a) except $U_{ij}=0$. (c) Instantaneous even-parity spectrum of the two-qubit two-mode quantum Rabi-Stark model during the adiabatic evolution process along $\vert \psi_{2s+}\rangle$ to generate $\vert W_2\rangle=\frac{1}{\sqrt{2}}\vert 10+01\rangle$ in $1.86\times 2\pi\omega^{-1}$.  $g=g_{11}=g_{12}=g_{21}=g_{22}$. (d) Instantaneous even-parity spectrum of the two-qubit single-mode quantum Rabi-Stark model during the adiabatic evolution process along $\vert \psi_{ds}\rangle$ to generate $\vert 1\rangle$ in $1.86\times2\pi\omega^{-1}$, with the same parameters as (c). $g=g_{11}=g_{12}$.}
\label{fig.2}
\end{figure}

For some cases we can much simplify the deduction: When $U_{ij}=U_j$, $\omega_i=\omega$ and $\frac{g_{ij}}{g_{i'j}}$ is independent of $j$, we can apply the same Bogoliubov transformation Eq. \eqref{bogo} to $H_S$, and obtain $H'_S=H'_{RS}+\sum_{i=2}^M(\omega+\sum_{j}U_{j}\sigma_{jz})b_i^\dagger b_i$ with
\begin{eqnarray}\label{hprs}
H'_{RS}&=&\sum_j\Delta_j\sigma_{jz}+(\omega+\sum_{j}U_{j}\sigma_{jz})b_1^\dagger b_1\nonumber\\ &&+\sum_{j}(\sum_ig_{ij}^2)^{\frac{1}{2}}\sigma_{jx} (b_1^\dagger+b_1).
\end{eqnarray}
So we only need to solve the single-mode model to obtain the full solution by changing $g_j$ into $(\sum_ig_{ij}^2)^{\frac{1}{2}}$ and $a_1$ into $b_1$ when $n_b=\sum_{i=2}^M n_{b_i}=0$. Otherwise, the qubit-part wavefunction will change because the free mode frequencies depend on $\sum_{j}U_{j}\sigma_{jz}$. However, $\hat{n}_{b_{i>1}}$ is still conserved, such that each solution has at least $n_{b}$ photons. 
For example, we can obtain $|\psi_{2s+}\rangle$ Eq. \eqref{dark1m} when $U_{ij}=U_j$ from the single mode solution \cite{sl}
\begin{equation}\small
\vert \psi_{ds}\rangle=\frac{1}{{\cal N}}[(\Delta_1-\Delta_2+U_1-U_2)\vert 0\uparrow\uparrow\rangle+g\vert 1( \downarrow\uparrow-\uparrow\downarrow)\rangle],
\end{equation} where $g=g_1=g_2$. If we carefully choose
$U_j$, then we can move many energy levels away from  $|\psi_{2s+}\rangle$. To make it more remarkable, we have taken the limit case $U_1+U_2=\omega$ to show the spectrum difference between the two-qubit two-mode Rabi-Stark model and corresponding Rabi model in Figs. \ref{fig.2} (a) and (b). In this case, the photon frequency is shifted down by $\omega$ for $\vert \downarrow\downarrow\rangle$, so all $\vert n_{1}n_{2}\downarrow\downarrow\rangle$ become degenerate when $g=0$. Numerical results show these infinite energy levels will lay below $E=-\omega$ as $g$ increases for even parity, which hardly affect the adiabatic speed along $|\psi_{2s+}\rangle$.
Interestingly, there is a dark state  $\vert \prod_j\downarrow_j\prod_i (\xi_i\to\infty)\rangle$ \cite{sl}
with constant eigenenergy $E=-\Delta_1-\Delta_2$ in the whole coupling regime, for $U_1+U_2=\omega$, as shown in Fig. \ref{fig.2} (a),(c), and (d), where$\vert\xi_i\to\infty\rangle=\sum_{n_i=0,1,2,\ldots}(-1)^{n_i}\sqrt{\frac{(2n_i)!}{2^{2n_i}(n_i!)^2}}\vert 2n_i\rangle$, a squeezed state with squeezing parameter tends infinity.

The enlarged energy gap between $\vert \psi_{2s+}\rangle$ and its closet energy levels can be used to accelerate the adiabatic evolution to generate $\vert W_M\psi_B\rangle$ from $\vert 0_M\uparrow\uparrow\rangle$. If we choose $U_1=U_2=\omega/2$, $\Delta_1(t=0)=\omega$, $\Delta_2(t=0)=0$, $g_1(t=0)=g_2(t=0)=0$, $\dot{\Delta}_2=-\dot{\Delta}_1=0.5/(1.86\times2\pi)\omega^{2}$, $\dot{g}_1=\dot{g}_2=0.7/(1.86\times2\pi)\omega^{2}$, then $\vert W_2\psi_B\rangle$ can be generated from $\vert 00\uparrow\uparrow\rangle$ in $1.86\times2\pi\omega^{-1}$ ($0.62$ ns )with fidelity reaching $99\%$ \cite{sl}. The instantaneous spectrum during the adiabatic evolution is shown in Fig. \ref{fig.2} (c). The blue energy levels $\vert \phi_K\rangle$ have $n_{b_2}=1$, which satisfy $\langle \phi_K|\dot{H}_S|\psi_{2s+}\rangle=0$, such that they will not affect the adiabatic evolution. The effective minimum energy gap reaches $0.55\omega$, ensuring the ultrafast $W$-state generation in a time proportional to the reverse of the resonator frequency, $f=\omega/2\pi$, through linear adiabatic passage. $\vert W_M\rangle$ can be generated with exact the same speed for different $M$ once $(\sum_ig_{i}^2)^{\frac{1}{2}}$ and other parameters are fixed, e.g., $\vert W_1\rangle=\vert 1\rangle$. The corresponding instantaneous spectrum is shown in Fig. \ref{fig.2} (d), which is exactly the same as  Fig. \ref{fig.2} (c), except the absence of energy levels $\vert\phi_K\rangle$ with $n_{b_{2}}\neq0$. 
\begin{figure}[tb]
\centering
\includegraphics[scale=0.39]{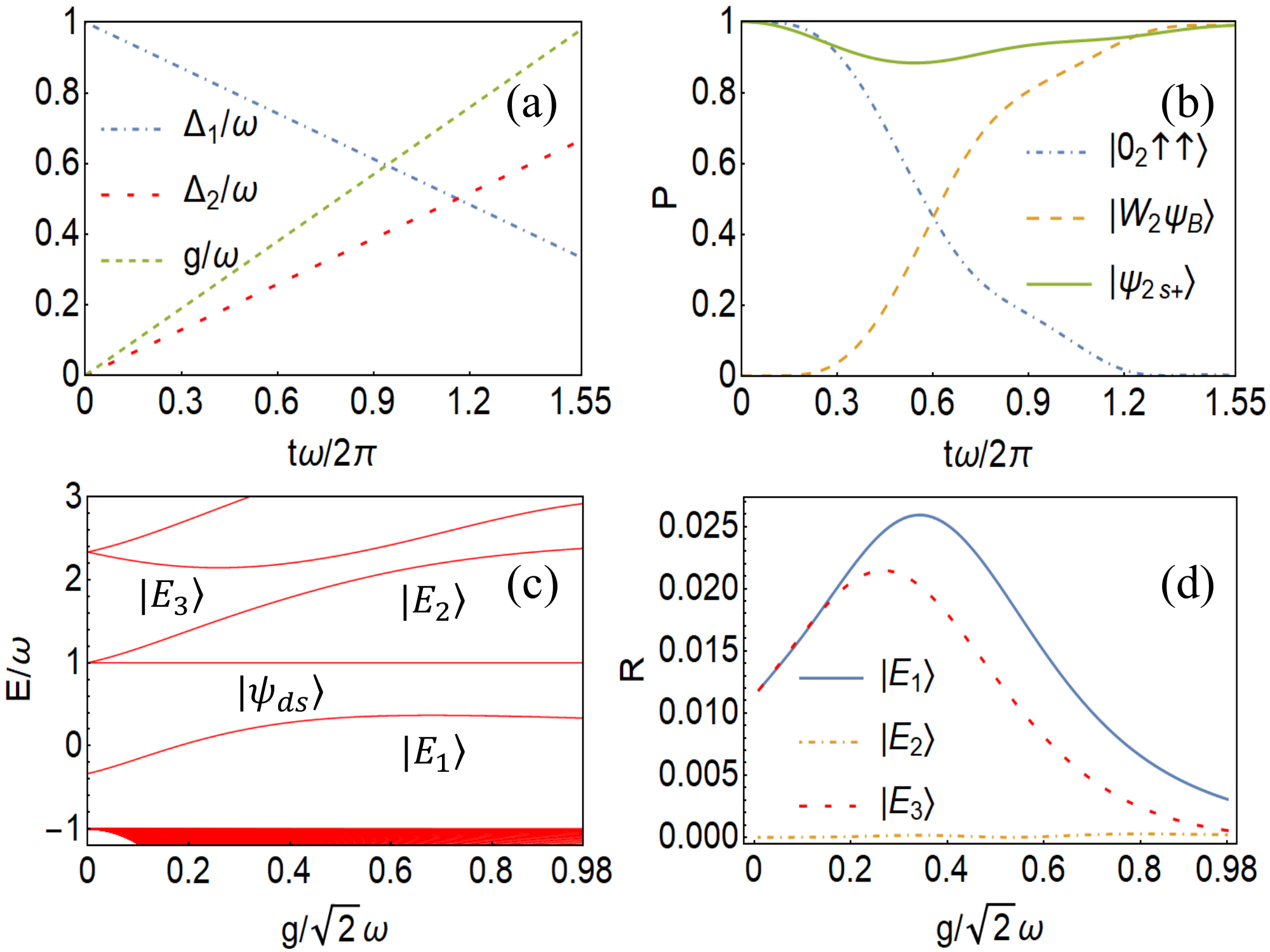}
\renewcommand\figurename{\textbf{FIG.}}
\caption[1]{(a) Evolution of parameters during the adiabatic process to generate $\vert W_2\rangle$. $g=g_{ij}$ for $i=1,2$ and $j=1,2$. $U_1=2\omega/3$. $U_2=\omega/3$. (b) Corresponding population of states. (c) Corresponding instantaneous spectrum of $H'_{RS}$. $g=(\sum_i g_{ij}^2)^{\frac{1}{2}}=\sqrt{2}g_{ij}$. (d) $R=\left|\frac{\langle E_{m}(t)|\dot{H}'_{RS}|\psi_{ds}(t)\rangle}{(E_m-\omega)^2}\right|$ for $\vert E_1\rangle$, $\vert E_2\rangle$ and $\vert E_3\rangle$ which are nearest to $\vert \psi_{ds+}\rangle$.
}
\label{figs4}
\end{figure}
This can be easily understood from $H'_{S}$ Eq. \eqref{hprs}.  Increasing $M$ will only add more energy levels $\vert\phi_K\rangle$ in the spectrum, which will not affect the adiabatic speed along $|\psi_{2s+}\rangle$, so arbitrary $\vert W_M\rangle$ can be generated with exactly the same speed. Different $U=U_1=U_2$ will bring different shift on the energy levels, such that different adiabatic speed. We search the   
the least generation time needed to reach above $99\%$ fidelity through linear adiabatic passages for different $U$, and find it approaches $1.86\times 2\pi\omega^{-1}$ for $U>0.4\omega$ \cite{sl}.  

However, even if a $\vert E_m\rangle$ with $n_{b_{i>1}}=0$ is very close to $\vert\psi_{2s+}\rangle$, the fast adiabatic evolution can be ensured by $\langle E_m=\omega+\delta|\dot{H}|\psi_{2s+}\rangle\propto f(\Delta_{1,2},U_{1,2},g_i)|\delta|/\sqrt{(\Delta_1+U_1-\Delta_2-U_2)^2+\delta^2}$ \cite{sl}, which can be further depressed if $U_1\neq U_2$ for fixed $\Delta_1-\Delta_2$. We find $\vert W_2\psi_B\rangle$ can be generated with fidelity $99\%$ in $1.55\times2\pi\omega^{-1}$ ($0.53$ ns) if $U_1=2/3\omega$, $U_2=1/3\omega$. The adiabatic trajectory is shown in Fig. \ref{figs4} (a), with corresponding population of states depicted in Fig. \ref{figs4} (b).
The instantaneous spectrum of $H'_{RS}$ Eq. \eqref{hprs} during the adiabatic process is shown in  Fig. \ref{figs4} (c), where $g=(\sum_i g_{ij}^2)^{\frac{1}{2}}$. We calculate $R=\left|\frac{\langle E_{m}(t)|\dot{H}'_{RS}|\psi_{ds}(t)\rangle}{(E_m-\omega)^2}\right|$ for $\vert E_1\rangle$, $\vert E_2\rangle$ and $\vert E_3\rangle$ which are nearest to $\vert \psi_{ds}\rangle$, and find  $R\ll1$, as shown in Fig. \ref{figs4} (d), so that the adiabatic condition Eq. \eqref{adia1} is satisfied. Interestingly, the closest one $\vert E_2\rangle$ has smallest and almost vanishing $R$, because its coefficients of  $\vert1_{b_1}\uparrow\downarrow\rangle$ and $\vert1_{b_1}\downarrow\uparrow\rangle$ are very small, and reduces to  $\vert3_{b_1}\downarrow\uparrow\rangle$ at $g=0$. So that the effective minimum energy gap $\omega-E_1$ reaches $0.63\omega$.

\begin{figure}[tb]
\centering
\includegraphics[scale=0.39]{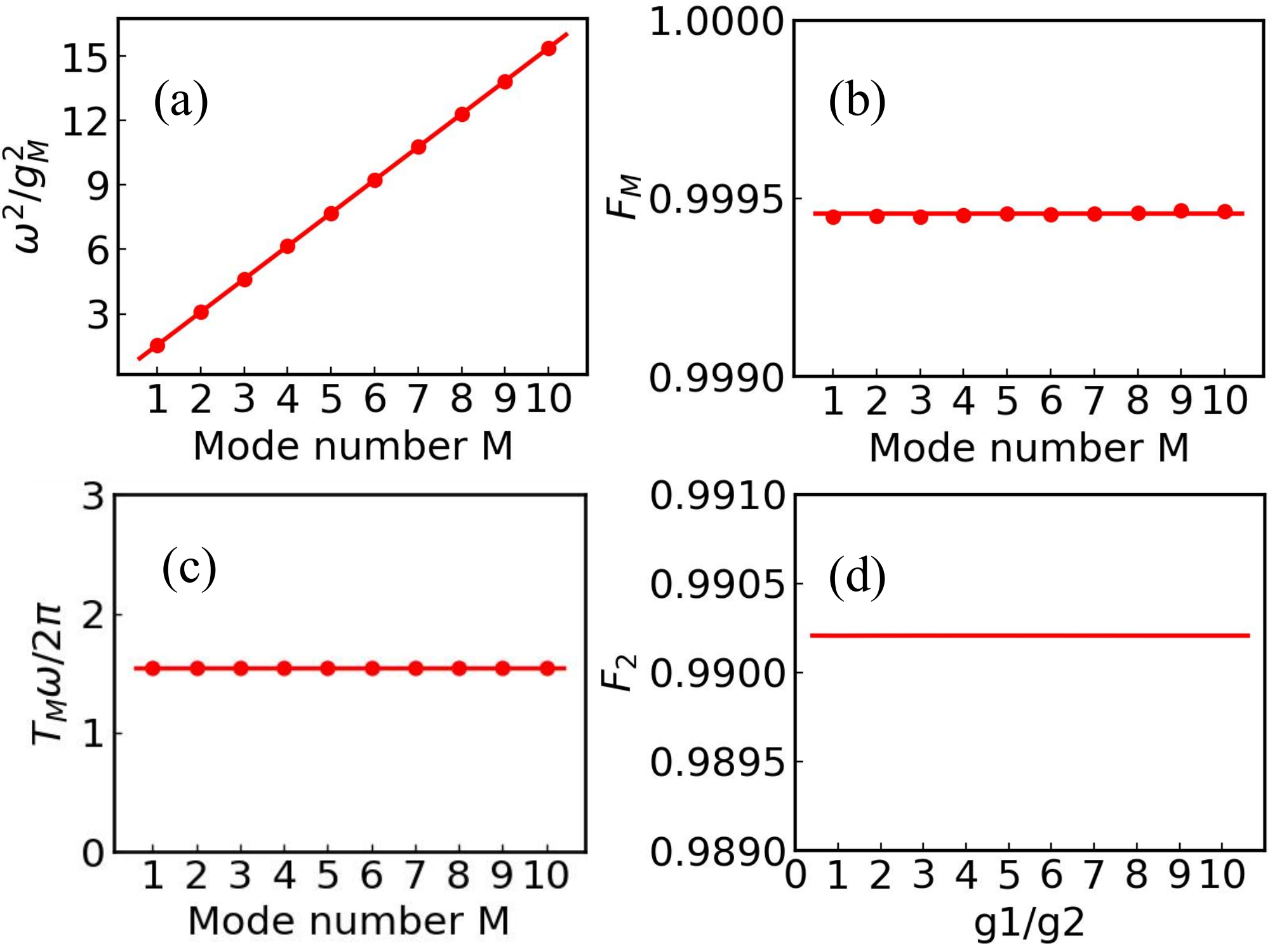}
\renewcommand\figurename{\textbf{FIG.}}
\caption[1]{Generation of $W$ states through adiabatic evolution. (a) We fix $\sum_{i=1}^M g_i^2=0.65\omega^2$ by choosing $g_i=g_M$ and 
$g_M=0.806\omega/\sqrt{M}$ with $M$ ranging from $1$ to $10$, so that we can generate $|W_M\rangle=\frac{1}{\sqrt{M}}\sum_{i=1}^M |0_1 0_2 \cdots 1_i 0_{i+1}\cdots 0_M\rangle$ with the same fidelity $F_M=99.95\%$ in $3.18\times 2\pi\omega^{-1}$ ($1.05$ ns), as shown in (b). (c) Generation time needed to reach $F_M=99\%$ of the above $|W_M\rangle$  for each $M$. (d) Fidelities to generate $\vert W_2\rangle=(g_1\vert 10\rangle+g_2\vert 01\rangle)/\sqrt{g_1^2+g_2^2}$ with $g_1/g_2$ ranging from 1 to 10 for fixed $g^2_1+g^2_2=0.98\omega^2$ within $T=1.86\times2\pi\omega^{-1}$ ($0.62$ ns). The linear adiabatic trajectories are shown in Ref. \cite{sl}. }
\label{fig.4}
\end{figure}

\begin{figure}[tb]
\centering
\includegraphics[scale=0.39]{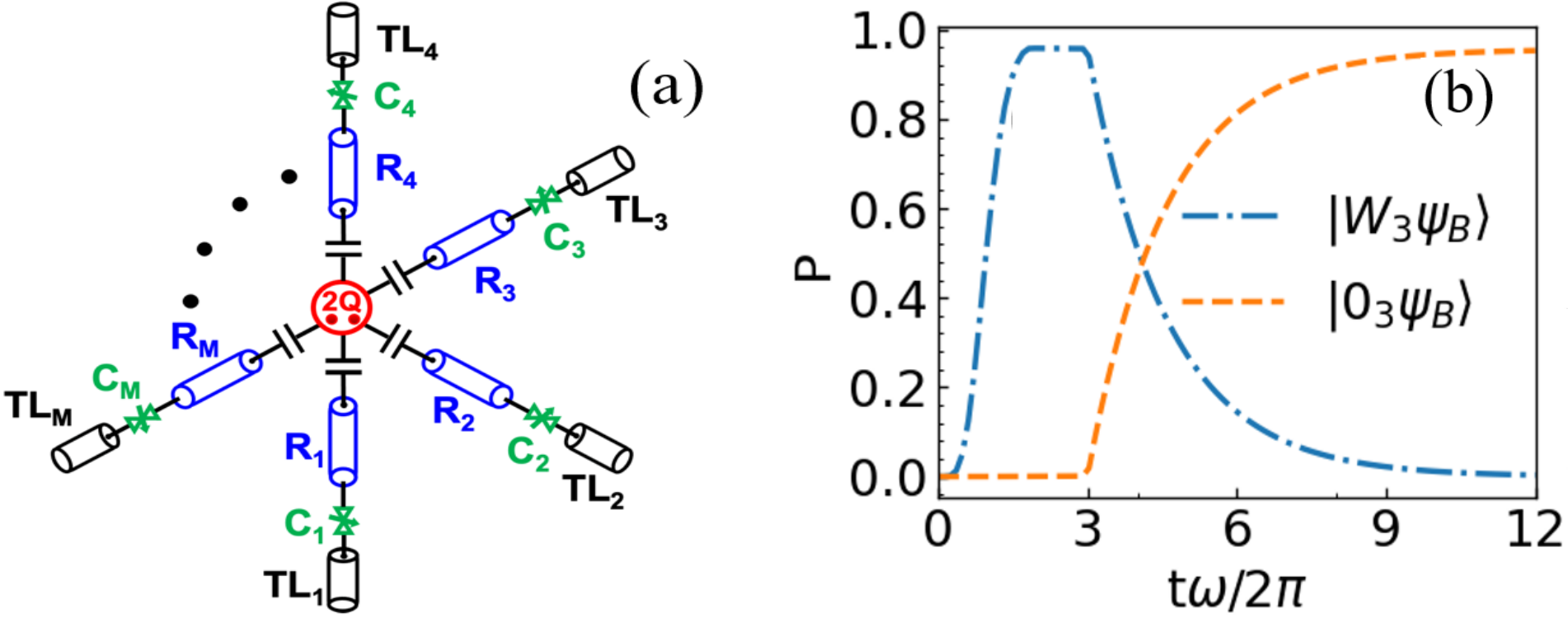}
\renewcommand\figurename{\textbf{FIG.}}
\caption[1]{(a) Schematic setup for the ultrafast generation of arbitrary $W$ states: Two qubits are coupled to $M$ resonators. Each
resonator is connected to a TL through a variable coupler C, so that the photon emission is controllable. (b) Numerical simulation for the catch and release of the $W$ state $|W_3\rangle=1/\sqrt{14}(|100\rangle+2|010\rangle+3|001\rangle)$ by solving the master equation \cite{sl}.
}
\label{fig.5}
\end{figure}

Now we show numerically arbitrary $W$ states can be ultrafast generated with exactly the same time and fidelity for fixed $\sum_{i=1}^M g_i^2$. To generate $|W_M\rangle=\frac{1}{M}\sum_{i=1}^M |0_1 0_2 \cdots 1_i 0_{i+1}\cdots 0_M\rangle$ for $M$ ranging from $1$ to $10$ with the same speed, we fix $\sum_{i=1}^M g_i^2=0.65\omega^2$, as shown in Fig. \ref{fig.4}(a). The fidelities of $\vert W_M\rangle$ all reach $99.95\%$ within $3.18\times2\pi\omega^{-1}$ ($1.06$ ns), as can be seen in Fig. \ref{fig.4}(b). Then we fix the fidelity to $99\%$, and find the least generation time needed for each $M$ is also the same, $1.55\times2\pi\omega^{-1}$ ($0.53$ ns), as shown in Fig. \ref{fig.4}(c). We also change the ratio between $g_1$ and $g_2$ while fix $\sum_{i=1}^2 g_i^2$ to generate different $\vert W_2\rangle=(g_1\vert 10\rangle+g_2\vert 01\rangle)/\sqrt{g_1^2+g_2^2} $, and find the fidelity is also the same within the same generation time, as shown in  Fig. \ref{fig.4} (d). The linear adiabatic trajectories for above cases are shown in \cite{sl}. Our scheme is robust under decoherence since the generation time is extremely small compared to decoherence time $\propto 10^{6}\omega^{-1}$ in circuit QED \cite{circuit}.

Our scheme to catch and release arbitrary $W$ states is shown in Fig. \ref{fig.5} (a). Two qubits are  coupled to  $M$ resonators
 with linear coupling  $g_i(\sigma_{1x}+\sigma_{2x})(a_i^\dag+a_i)$ ($i=1,2,3\ldots M$) and nonlinear 
coupling $(U_1\sigma_{1z}+U_2\sigma_{2z})a^\dag_ia_i$. Each resonator is connected to 
a transmission line (TL) through a variable coupler $C$, so that the photon emission is controllable \cite{yiyin}. $|W_3\rangle$ is generated in $t=1.86\times2\pi\omega^{-1}$, and then we open the external dissipation channel at $t=3\times2\pi\omega^{-1}$ to release them into the transmission lines  \cite{sl}, as simulated in Fig.  \ref{fig.5} (b).

\emph{Conclusion.--}The solutions to the QRM can not only have photon-number upper bound when generalized to the multiqubit case, but also have lower bound when generalized to the multimode case. Their peculiarities and the reach of ultrastrong coupling ($g\gtrsim0.1\omega$) make it possible to realize a simple linear ultrafast adiabatic passage to generate arbitrary $W$ states with the same speed, since it is symmetry protected and the effective minimum energy gap reaches $0.63\omega$. 
Our results show although generally infinite photons are excited by the ultrastrong coupling, it is possible to use its photon-number bounded solutions to realize ultrafast and deterministic quantum information protocols through adiabatic evolution.

\emph{Acknowledgements.--}
This work was supported
by the Natural Science Foundation of Hunan Province,
China (Grant No. 2022JJ30556, 2023JJ30596, 2023JJ30588), Education Department of Hunan Province in China (Grant No. 23A0135, 21B0138), the National Natural Science
Foundation of China (Grant Nos. 12035011, 11704320).

\newcommand{\red}[1]{\textcolor{red}{#1}}
\newcommand{\blue}[1]{\textcolor{blue}{#1}}
\newcommand{\green}[1]{\textcolor{green}{#1}}
\newcommand{\magenta}[1]{\textcolor{magenta}{#1}}
\newcommand{\cyan}[1]{\textcolor{cyan}{#1}}
\newcommand{\brown}[1]{\textcolor{brown}{#1}}
\newcommand{\ket}[1]{\vert #1 \rangle}
\newcommand{\bra}[1]{\langle #1 \vert}
\newcommand{\ketbra}[2]{\vert #1 \rangle \langle #2 \vert}
\newcommand{\braket}[2]{\langle #1 \vert #2 \rangle}
\newcommand{\eg}{{\it{e.g.}}}
\newcommand{\ie}{{\it{i.e.}}}

\widetext
\begin{center}
\textbf{ \large Supplemental Material: \\ Ultrafast adiabatic passages in ultrastrongly coupled light-matter systems}
\end{center}

\renewcommand{\thesection}{S\arabic{section}}
\renewcommand{\thesubsection}{\Alph{subsection}}
\renewcommand{\thesubsubsection}{\alph\arabic{subsubsection}}
\renewcommand{\theequation}{S\arabic{equation}}
\renewcommand{\thefigure}{S\arabic{figure}}
\renewcommand{\thetable}{S\arabic{table}}
\setcounter{equation}{0}
\setcounter{figure}{0}

~\\

This supplemental material contains five parts: (1) Proof of  $\langle\psi_{E=\omega+\delta}|\dot{H}|\psi_{2+}\rangle\propto f(\Delta_{1,2},g_i)|\delta|/\sqrt{(\Delta_1-\Delta_2)^2+\delta^2}$. (2) Method to obtain the one-photon dark-sate solutions of the multi-qubit $M$-mode quantum Rabi-Stark model. (3) Method to obtain the dark-state solution $\vert \prod_j\downarrow_j\prod_i (\xi_i\to\infty)\rangle$. (4) Proof of $\langle\psi_{E=\omega+\delta}|\dot{H}|\psi_{2s+}\rangle\propto f(\Delta_{1,2},U_{1,2},g_i)|\delta|/\sqrt{(\Delta_1-\Delta_2+U_1-U_2)^2+\delta^2}$. (5) The least generation time needed to reach above $99\%$ fidelity for different $U$. (6) Adiabatic trajectories to generate different $W$ states discussed in the main text. (7) Discussion of the effect of the environment and catch and release of the $W$ states.

\section{Proof of  $\langle\psi_{E=\omega+\delta}|\dot{H}|\psi_{2+}\rangle\propto f(\Delta_{1,2},g_i)|\delta|/\sqrt{(\Delta_1-\Delta_2)^2+\delta^2}$}
According to the adiabatic theorem \cite{amin,aiu,bda}, the system will evolve along $\vert\psi_{2+}\rangle$ adiabatically if
\begin{equation}\label{adia}
\left|\frac{\langle E_{m}(t)|\dot{H}|\psi_{2+}(t)\rangle}{(E_m-\omega)^2}\right|\ll1 ~~~t\in[0,T],
\end{equation}
since $\vert\psi_{2+}\rangle$ has constant eigenenergy $\omega$.
However, as can be seen in Fig. 1 of the main text, there is an energy level very close to $|\psi_{2+}\rangle$, which seems require a slow change of $H$. But actually, arbitrary $\vert
W_M\psi_B\rangle$ can be generated from $\vert 0\uparrow\uparrow\rangle$ in $11\times 2\pi\omega^{-1}$ with fidelity $99.17\%$ through adiabatic evolution along $\vert\psi_{2+}\rangle$. This is because $\langle\psi_{E=\omega+\delta}|\dot{H}|\psi_{2+}\rangle\propto|\delta|/\sqrt{(\Delta_1-\Delta_2)^2+\delta^2}$, which decreases as  $\vert E_{m}\rangle$ goes closer to $|\psi_{2+}\rangle$.

As shown in Eq. (9) of the main text,
\begin{eqnarray}\label{hdot}
\dot{H}=\dot{\Delta}_1(\sigma_{1z}-\sigma_{2z})+(\sum_{i}g_{i}^2)^{\frac{1}{2}}(b_1+b_1^\dagger)(\sigma_{1x}+\sigma_{2x})\dot{g_i}/g_i,
\end{eqnarray}
and $|\psi_{2+}\rangle$ reads (Eq. (6) of the main text)
\begin{equation}\label{psi2+}
\vert \psi_{2+}\rangle=\frac{1}{\cal{N}}[(\Delta_1-\Delta_2)\vert 0_M\uparrow\uparrow\rangle+(\sum_{i}g_{i}^2)^{\frac{1}{2}}\vert 1_{b_1}(\downarrow\uparrow-\uparrow\downarrow)\rangle].
\end{equation} 
So 
\begin{equation}\label{co1}
\dot{H}|\psi_{2+}\rangle=\frac{(\sum_{i}g_{i}^2)^{\frac{1}{2}}}{{\cal
N}}(|1_{b_1}\downarrow\uparrow\rangle+
 |1_{b_1}\uparrow\downarrow\rangle)\left[(\Delta_1-\Delta_2)\dot{g_i}/g_i-2\dot{\Delta}_1 
\right].
\end{equation}
It can be seen that although $\vert E_m\rangle$ consists of infinite basis, only $\vert1_{b_1}\uparrow\downarrow\rangle$ and $\vert1_{b_1}\downarrow\uparrow\rangle$ will affect the adiabatic speed considering Eqs. \eqref{adia} and \eqref{co1}.
According to $\langle1_{b_1}\downarrow\uparrow-1_{b_1}\uparrow\downarrow|H-\omega-\delta|\psi_{E=\omega+\delta}\rangle=0$, we obtain
\begin{eqnarray}
(\Delta_2-\Delta_1-\delta)\langle 1_{b_1}\downarrow\uparrow|\psi_{E=\omega+\delta}\rangle=(\Delta_1-\Delta_2-\delta)\langle 1_{b_1}\uparrow\downarrow|\psi_{E=\omega+\delta}\rangle.
\end{eqnarray}
Therefore,
\begin{eqnarray}\label{coe}
\langle\psi_{E=\omega+\delta}|\dot{H}|\psi_{2+}\rangle\propto f(\Delta_{1,2},g_{i})\frac{|\delta|}{\sqrt{(\Delta_1-\Delta_2)^2+\delta^2}}
\end{eqnarray}
considering Eq. \eqref{co1}.
\section{One-photon dark-sate solutions of the multiqubit $M$-mode quantum Rabi-Stark model}
The multiqubit $M$-mode quantum Rabi-Stark model reads
\begin{eqnarray}\label{hs}
H_S=\sum_{j=2}^{N}\Delta_j\sigma_{jz}+\sum_{i=1}^{M}(\omega_i+\sum_{j=2}^N U_{ij}\sigma_{jz})a_i^\dagger a_i +\sum_{i=1,j=2}^{M,N}g_{ij}(a_i^\dagger+a_i)\sigma_{jx}.
\end{eqnarray}
It has a $\mathbb{Z}_2$ symmetry with generator $R=\exp [i\pi \sum_{i=1}^M a_i^\dag a_i]\Pi_j\sigma_{jz}$. According to its eigenvalues p, we divide the Hilbert space into 
\begin{eqnarray}
&(|0_M, \psi_{+}\rangle)\leftrightarrow (|1_M,
\psi_{-}\rangle)\leftrightarrow (|2_M,
\psi_{+}\rangle) \cdots (p=1)\\ 
&(|0_M, \psi_{-}\rangle)\leftrightarrow(|1_M,
\psi_{+}\rangle)\leftrightarrow (|2_M,
\psi_{-}\rangle) \cdots (p=-1)
\end{eqnarray}
where $|k_M\rangle$ represents all $M$-mode states with $k$ photons. $|\psi_{\pm}\rangle$ represents all qubit states with eigenvalues of $\Pi_j\sigma_{jz}$ being $\pm1$. Supposing there are eigenstates with at most one photon $|\psi_{s\pm} \rangle=\vec{c}^\pm_{0,M}|0_M, \psi_{\pm}\rangle+\vec{c}^\pm_{1,M}|1_M,\psi_{\mp}\rangle$, where $\vec{c}_{k,M}^\pm$ is the coefficient vector. Then the eigenenergy equation reads
\begin{eqnarray}\label{1pho}
\left(
  \begin{array}{cc}
   D_{0}^\pm-E &O^\dag_{~0} \\
    O_{~0} & D_{~1}^\pm-E  \\
      0& O_{~1} \\
  \end{array}
\right)\left(
  \begin{array}{c}
   \vec{c}^\pm_{0_M} \\
   \vec{c}^\pm_{1_M} \\
  \end{array}
\right)=0 ,
\end{eqnarray}
where $D_{k}^\pm$ is a $2^{N-1}C_{M+k-1}^{k}\times 2^{N-1}C_{M+k-1}^{k}$
matrix, and $O_{k}$ is a  $2^{N-1}C_{M+k}^{k+1}\times 2^{N-1}C_{M+k-1}^{k}$ matrix, which takes the same form for even and odd parities. Obviously there are more equations than variables, so normally there is no nontrivial solution. However, when parameters in the coefficient matrix of Eq. \eqref{1pho} satisfy certain condition, the matrix will have less nonzero rows than columns after elementary row transformations, such that nontrivial solution exists.  

Let us start with the two-qubit single mode case. Now $|\psi_{s+} \rangle=c_{0} \vert 0,\uparrow,\uparrow\rangle+c_{1} \vert 0,\downarrow,\downarrow\rangle+c_{2} \vert 1,\downarrow,\uparrow\rangle+c_{3} \vert 1,\uparrow,\downarrow\rangle$ for even parity, and Eq. \eqref{1pho} reads
\begin{eqnarray}
\small\label{eigcq}\nonumber
  \left(\begin{array}{cccccc}
  \Delta_1+\Delta_2-E &0&g_{11} & g_{21} \\
   0&-\Delta_1-\Delta_2-E &g_{21}&g_{11} \\
   g_{11} & g_{21}& \omega-\Delta_1+\Delta_2-U_{11}+U_{21}-E&0\\
    g_{21} & g_{11}&0& \omega+\Delta_1-\Delta_2+U_{11}-U_{21}-E \\
           0&0&\sqrt{2}g_{11} & \sqrt{2}g_{21} \\
            0&0&\sqrt{2}g_{21} & \sqrt{2}g_{11} \\
  \end{array}\right) 
  \left(\begin{array}{c}
  c_{0} \\
   c_{1} \\
   c_{2}\\
   c_{3}\\
  \end{array}\right)=0 .
\end{eqnarray}
We apply elementary row transformations to the $6\times4$ matrix and find the nonzero rows can be less than the columns when $\Delta_1+\Delta_2=\omega=E$ and $g_{11}=g_{21}=g$, with the matrix transformed into
\begin{equation} \label{eigen2}
\begin{pmatrix}
   g &0 & -\Delta_1+\Delta_2-U_{11}+U_{21}&0\\
   0 & 1&0&0 \\
  0&0&1&1\\
   0&0 &0&0\\
    0&0&0&0\\
    0&0&0&0\\
\end{pmatrix},
\end{equation}
which means nontrival solution
\begin{equation}
\vert \psi_{ds}\rangle=(\Delta_1-\Delta_2+U_{11}-U_{21})\vert 0\uparrow\uparrow\rangle+g\vert 1\rangle(\vert \downarrow\uparrow-\uparrow\downarrow\rangle)
\end{equation}
exist for arbitrary $g$ with constant energy, corresponding to a horizontal line in the spectrum, which is a special dark state.

Next we consider the two-qubit two-mode case, where the coefficient matrix in linear Eq. \eqref{1pho} with even parity becomes
\begin{eqnarray}
\nonumber
\scriptsize
\setlength{\arraycolsep}{-4pt}
  \begin{pmatrix}
   \Delta_1+\Delta_2-E&0&g_{12}&g_{11}&g_{22}&g_{21}\\
     0&-\Delta_1-\Delta_2-E&g_{11}&g_{12}&g_{21}&g_{22}\\
     g_{12}&g_{11}&\omega_1+\Delta_1-\Delta_2+U_{11}-U_{12}-E&0&0&0\\
     g_{11}&g_{12}&0&\omega_1-\Delta_1+\Delta_2-U_{11}+U_{12}-E&0&0\\
     g_{22}&g_{21}&0&0&\omega_2+\Delta_1-\Delta_2+U_{21}-U_{22}-E&0\\
     g_{21}&g_{22}&0&0&0&\omega_2-\Delta_1+\Delta_2-U_{21}+U_{22}-E\\
    0&0&\sqrt2g_{12}&\sqrt2g_{11}&0&0\\
    0&0&\sqrt2g_{11}&\sqrt2g_{12}&0&0\\
    0&0&g_{12}&g_{21}&g_{12}&g_{11}\\
    0&0&g_{21}&g_{22}&g_{11}&g_{12}\\
    0&0&0&0&\sqrt2g_{22}&\sqrt2g_{21}\\
    0&0&0&0&\sqrt2g_{21}&\sqrt2g_{22}\\   
    \end{pmatrix}.
\end{eqnarray}
Eq. \eqref{1pho} has nontrivial solution if the above matrix has less nonzero rows than columns after elementary row transformation, which can be satisfied when $\omega_1=\omega_2=E=\Delta_1+\Delta_2$, $g_{11}=g_{12}=g_1$ and $g_{21}=g_{22}=g_2$. Such solution reads
\begin{equation}
\begin{aligned}
|\psi_{s+}\rangle =\frac{1}{{\cal N}}[|00\uparrow\uparrow\rangle+(\frac{g_1}{\Delta_1-\Delta_2+U_{11}-U_{12}}|10\rangle+
	\frac{g_2}{\Delta_1-\Delta_2+U_{21}-U_{22}}|01\rangle)(|\downarrow\uparrow\rangle-|\uparrow\downarrow\rangle)]
\end{aligned}
\end{equation}
which exist in the whole coupling regime with constant eigenenergy $E=\omega$.
Following similar deduction, we find a general one-photon quasi-exact solution with even parity for the two-qubit $M$-mode quantum Rabi-Stark model
\begin{equation}\label{dark1}
	\vert\psi_{2s+}\rangle =(1/{{\cal N}})\left[|0_M\uparrow\uparrow\rangle+|W'_M(\downarrow\uparrow-\uparrow\downarrow)\rangle\right]
\end{equation}
under the condition $\omega_i=E^+=\Delta_1+\Delta_2$ and $g_{i1}=g_{i2}=g_i$ for i from 1 to M, where $|W'_M\rangle=\frac{g_1}{\Delta_1-\Delta_2+U_{11}-U_{12}}|100\cdots0\rangle+\frac{g_2}{\Delta_1-\Delta_2+U_{21}-U_{22}}|010\cdots0\rangle+\cdots+\frac{g_M}{\Delta_1-\Delta_2+U_{M1}-U_{M2}}|000\cdots1\rangle$. Such one-photon solutions for odd parity read
\begin{eqnarray}
&|\psi_{2s,a-}\rangle =\frac{1}{{\cal N'}}\left[|0_M\uparrow\downarrow\rangle+|W_M'(\downarrow\downarrow-\uparrow\uparrow)\rangle\right]\qquad\\
&|\psi_{2s,b-}\rangle =\frac{1}{{\cal N'}}\left[|0_M\downarrow\uparrow\rangle+|W_M'(\downarrow\downarrow-\uparrow\uparrow)\rangle\right]\qquad
\end{eqnarray}
with the condition $\omega_i=E^-=\Delta_1-\Delta_2$ and $\omega_i=E^-=\Delta_2-\Delta_1$ respectively, and $g_{i1}=g_{i2}=g_i$. Here $|W'_M\rangle=\frac{g_1}{\Delta_1+\Delta_2+U_{11}+U_{12}}|100\cdots0\rangle+\frac{g_2}{\Delta_1+\Delta_2+U_{21}+U_{22}}|010\cdots0\rangle+\cdots+\frac{g_M}{\Delta_1+\Delta_2+U_{M1}+U_{M2}}|000\cdots1\rangle$.

For the three-qubit and $M$-mode case, we find a one-photon solution for odd parity 
\begin{eqnarray}
|\psi_{3s-}\rangle=&&|W_M\rangle(|\uparrow\downarrow\downarrow\rangle-|\downarrow\uparrow\downarrow\rangle
-|\downarrow\downarrow\uparrow\rangle+|\uparrow\uparrow\uparrow\rangle)
+\frac{\omega g_{13}-g_{12}U_{11}+g_{11}U_{12}}{g_{12}}|0_M\uparrow\uparrow\downarrow\rangle\nonumber \\ &&+\frac{\omega g_{12}-g_{13}U_{11}+g_{11}U_{13}}{g_{13}}|0_M\uparrow\downarrow\uparrow\rangle
-\frac{g_{11}(\omega g_{11}+g_{13}U_{12}+g_{12}U_{13})}{g_{12}g_{13}}|0_M\downarrow\uparrow\uparrow\rangle ,
\end{eqnarray}
with the condition $\Delta_j=\omega_i=\omega=E^{-}$, $g_{i1}=g_{i2}+g_{i3}$, $g_{i2}/g_{i3}=g_{12}/g_{13}$ and $U_{i1}=U_{11}$, $U_{i2}=U_{12}$, $U_{i3}=U_{13}$, where $|W_M\rangle= g_{11}|100\ldots0\rangle+g_{21}|010\ldots0\rangle+\dots+g_{M1}|000\ldots1\rangle$.

Therefore, the one-photon solution with constant eigenenergy $E=\omega$ for the N-qubit and M-mode case reads
\begin{eqnarray}
&&|\psi_{N}\rangle=|\psi_{2s}\rangle\otimes(|\psi_B\rangle)^{(N-2)/2},\ \ \ \ N=2,4,6,\dots,\\
&&|\psi_{N}\rangle=|\psi_{3s}\rangle\otimes(|\psi_B\rangle)^{(N-3)/2},\ \ \ \ N=3,5,7,\dots,
\end{eqnarray}where $|\psi_B\rangle=1/\sqrt{2} (|\downarrow\uparrow\rangle-|\uparrow\downarrow \rangle)$, a two-qubit singlet Bell state and $|\psi_{2s,3s}\rangle$ are obtained as above.

\section{Method to obtain the dark-state solution $\vert \prod_j\downarrow_j\prod_i (\xi_i\to\infty)\rangle$ }
There are dark-state solutions for the multiqubit $M$-mode Rabi-Stark model $H_S$ (Eq. \eqref{hs}) whose qubit part consists of only $\vert\prod_j\downarrow_j\rangle$ when $\omega_{i}=\sum_{j}U_{ij}$. The reason is as follows: $\omega_i a_i^\dag a_i$ are cancelled by $\sum_j U_{ij}\sigma_{jz}a_i^\dag a_i$ for $\vert \prod_j \downarrow_j\rangle$, so $H_{S}$ represents a zero-frequency harmonic oscillator, and all $ f(a_i^\dag)\vert0_M\rangle\vert \prod_j \downarrow_j\rangle$ become degenerate when $g=0$. If $g\neq0$, the effective Hamiltonian becomes $\sum_j\Delta_j\sigma_{jz}+\sum g_{ij}(a_i^\dagger+a_i)\sigma_{jx}$ for $ f(a_i^\dag)\vert0_M\rangle\vert \prod_j \downarrow_j\rangle$, which seems to have no solution. However, a squeezed state $\vert \xi_i\rangle$ 
 \begin{equation}
\vert\xi_i\rangle=\sum_{n=0,1,2,\ldots}(-1)^n\sqrt{\frac{(2n)!}{2^{2n}(n!)^2}}\vert 2n\rangle.
\end{equation} 
is the eigenstate of $(a_i+a_i^\dag)$ with zero eigenvalue when $\vert\xi_i\vert\rightarrow\infty$ and squzeeing angle $\theta=0$. The only term left in the effective Hamiltonian is $\sum_j\Delta_j\sigma_{jz}$, so there is a dark state $\vert\psi_{D}\rangle=\vert \prod_j\downarrow_j\prod_i (\xi_i\to\infty)\rangle$
with constant eigenenergy $E=-\sum_j\Delta_j$ in the whole coupling regime, as shown in Fig. (2) of the main text. This state is dark not because the photon number is bounded, but the qubit states are restricted to only $\vert \prod_j\downarrow_j\rangle$, although $H_{S}$ connects it to other states. The most distinct feature is this state is totally a product state, leading to entanglement death, different from other special dark states.

\section{Proof of $\langle\psi_{E=\omega+\delta}|\dot{H}|\psi_{2s+}\rangle\propto f(\Delta_{1,2},U_{1,2},g_i)|\delta|/\sqrt{(\Delta_1-\Delta_2+U_1-U_2)^2+\delta^2}$}
As shown in Fig. 3 (b) of the main text,  $\vert W_2\rangle$ can be generated in $1.55\times 2\pi\omega^{-1}$ through the adiabatic evolution along $\vert \psi_{2s+}\rangle$. This is because the peculiarity of $\vert \psi_{2s+}\rangle$, including $\langle E=\omega+\delta|\dot{H}|\psi_{2+}\rangle\propto|\delta|/\sqrt{(\Delta_1-\Delta_2+U_1-U_2)^2+\delta^2}$.

Similar to Eq. \eqref{hdot},
\begin{eqnarray}
\dot{H}_S=(\dot{\Delta}_1+\sum_i \dot{U}_1 b_i^\dag b_i)\sigma_{1z}+(-\dot{\Delta}_1+\sum_i \dot{U}_2 b_i^\dag b_i)\sigma_{2z}+(\sum_{i}g_{i}^2)^{\frac{1}{2}}(b_1+b_1^\dagger)(\sigma_{1x}+\sigma_{2x})\dot{g_i}/g_i,
\end{eqnarray}
for $U_{i1}=U_1$ and $U_{i2}=U_2$, and $|\psi_{2s+}\rangle$ reads (Eq. (11) of the main text)
\begin{equation}\label{psi2+s}
\vert \psi_{2s+}\rangle=\frac{1}{\cal{N}}[(\Delta_1+U_1-\Delta_2-U_2)\vert 0_M\uparrow\uparrow\rangle+(\sum_{i}g_{i}^2)^{\frac{1}{2}}\vert 1_{b_1}(\downarrow\uparrow-\uparrow\downarrow)\rangle].
\end{equation} 
So 
\begin{equation}\label{co1s}
\dot{H}_S|\psi_{2s+}\rangle=\frac{(\sum_{i}g_{i}^2)^{\frac{1}{2}}}{{\cal
N}}(|1_{b_1}\downarrow\uparrow\rangle+
 |1_{b_1}\uparrow\downarrow\rangle)\left[(\Delta_1+U_1-\Delta_2-U_2)\dot{g_i}/g_i-2\dot{\Delta}_1-\dot{U}_1+\dot{U}_2 
\right].
\end{equation}
It can be seen that although other energy levels $\vert E=\omega+\delta\rangle$ consists of infinite basis, only $\vert1_{b_1}\uparrow\downarrow\rangle$ and $\vert1_{b_1}\downarrow\uparrow\rangle$ will affect the adiabatic speed considering Eqs. \eqref{adia} and \eqref{co1s}.
According to $\langle1_{b_1}\downarrow\uparrow-1_{b_1}\uparrow\downarrow|H-\omega-\delta|E=\omega+\delta\rangle=0$, we obtain
\begin{eqnarray}
(\Delta_2+U_2-\Delta_1-U_1-\delta)\langle 1_{b_1}\downarrow\uparrow|E=\omega+\delta\rangle=(\Delta_1+U_1-\Delta_2-U_2-\delta)\langle 1_{b_1}\uparrow\downarrow|E=\omega+\delta\rangle.
\end{eqnarray}
Considering Eq. \eqref{co1s}, we obtain
\begin{eqnarray}\label{coe}
\langle\psi_{E=\omega+\delta}|\dot{H}|\psi_{2s+}\rangle\propto f(\Delta_{1,2},U_{1,2},g_i)|\delta|/\sqrt{(\Delta_1-\Delta_2+U_1-U_2)^2+\delta^2},
\end{eqnarray}
which can be further depressed if $U_1\neq U_2$.

\section{Time needed to generate $\vert W_2\psi_B\rangle$ with  fidelity $99\%$ for different $U$}
\begin{figure}[h]
\centering
\resizebox{0.4\columnwidth}{!}{
\includegraphics{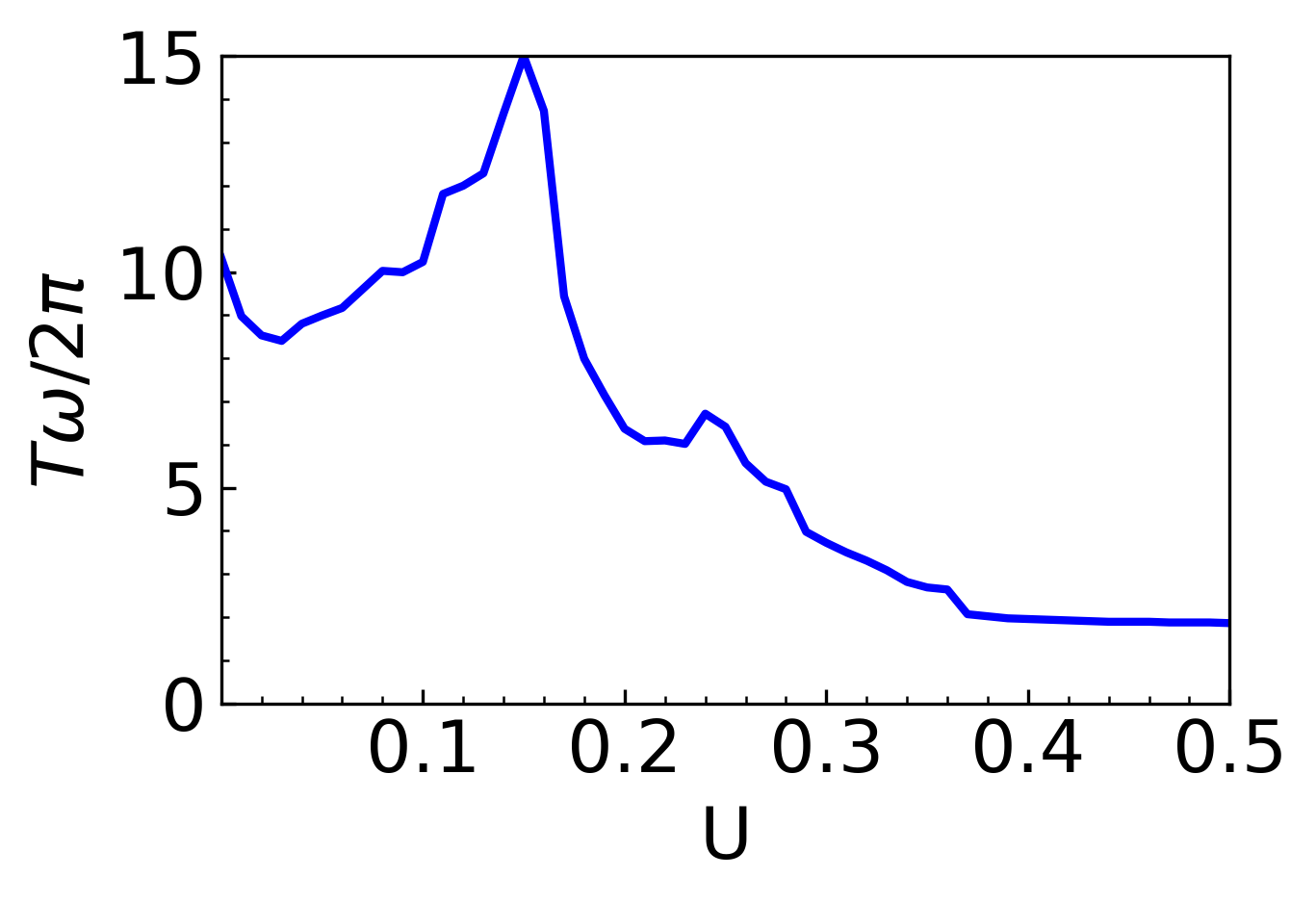}
}
\renewcommand\figurename{\textbf{FIG.}}
\caption[1]{The least generation time needed to generate $\vert W_2\psi_B\rangle$ with  fidelity above $99\%$ for different $U$.}
\label{figu}
\end{figure}

$\vert W_2\psi_B\rangle$ can be generated from $\vert 0_2\uparrow\uparrow\rangle$ through adiabatic evolution along $\vert \psi_{2s+}\rangle$ in a time $T$. We choose $U_1=U_2=U$, $g_1=g_2=g$. $\Delta_{1,2}(t)$ evolve linearly from $\omega$ and $0$ to $\omega/2$, respectively, while $g(t)$ evolves linearly from and $0$ to $g_{max}$. We search the least $T$ needed to reach above $99\%$ fidelity for $g_{max}$ in $(0,1.2\omega)$ and $U$ in $(0,0.5\omega)$. Different $U$ will bring
different shift on the energy levels, such that different adiabatic speed and least generation time $T$, as shown in Fig. \ref{figu}. It can be seen that $U = 0.5\omega$ is the
best choice with $T = 1.86\times2\pi\omega^{-1}$, which is almost unchanged
for $U > 0.37$. 

\section{Adiabatic trajectories to generate different $W$ states mentioned in the main text}
\begin{figure}[h]
\centering
\resizebox{1\columnwidth}{!}{
\includegraphics{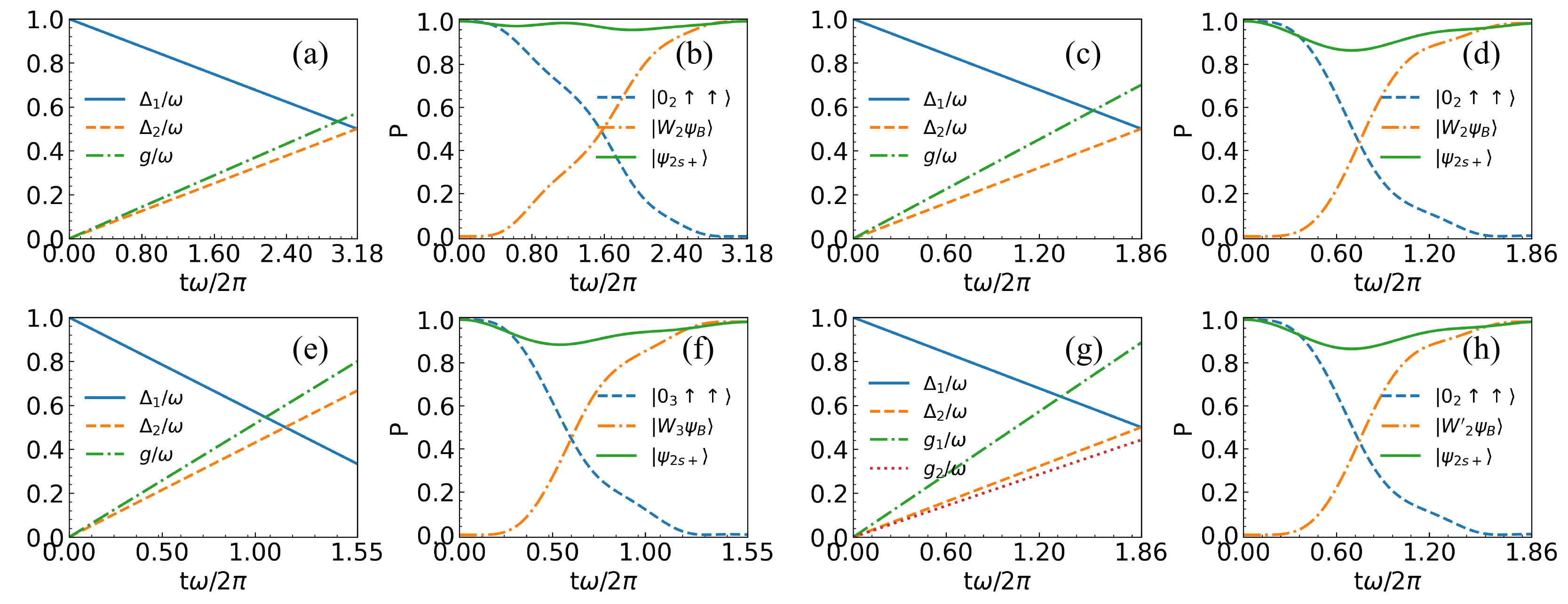}
}
\renewcommand\figurename{\textbf{FIG.}}
\caption[1]{(a) Adiabatic trajectory used to vary the parameters for generating $\vert W_2\psi_B\rangle$ in $3.18\times2\pi\omega^{-1}$ with fidelity $99.95\%$. $\vert W_2\rangle=\frac{1}{\sqrt{2}}(\vert 10\rangle+\vert 01\rangle)$. $g_1=g_2=g$. (b) Population of states when parameters evolve as in (a). (c) Adiabatic trajectory used to vary the parameters for generating $\vert W_2\psi_B\rangle$ in $1.86\times2\pi\omega^{-1}$ with fidelity $99\%$. (d) Population of states when parameters evolve as in (c). (e) Adiabatic trajectory used to vary the parameters for generating $\vert W_3\psi_B\rangle$ in $1.55\times2\pi\omega^{-1}$ with fidelity $99\%$. $\vert W_3\rangle=\frac{1}{\sqrt{3}}(\vert 100\rangle+\vert 010\rangle+\vert 001\rangle)$. $g_1=g_2=g_3=g$. (f) Population of states when parameters evolve as in (e). (g) Adiabatic trajectory used to vary the parameters for generating $\vert W'_2\psi_B\rangle$ in $1.86\times2\pi\omega^{-1}$ with fidelity $99\%$. $\vert W'_2\rangle=\frac{1}{\sqrt{g_1^2+g_2^2}}(g_1\vert 10\rangle+g_2\vert01\rangle)$. $g_1=2g_2$. (h) Population of states when parameters evolve as in (g).}
\label{fig1}
\end{figure}
As described in Fig. 2 (c) of the main text, $\vert W_2 \psi_B\rangle$ can be generated in $1.86\times2\pi\omega^{-1}$ with fidelity $99\%$ through adiabatic evolution along $\vert \psi_{2s+}\rangle$. Here we show its adiabatic trajectory in Fig. \ref{fig1} (c) and (d), where the state population is obtained by solving the Schr\"{o}dinger equation.  Meanwhile, as shown in Fig. 4 (b) of the main text, $\vert W_M \psi_B\rangle$ can be generated in $3.18\times2\pi\omega^{-1}$ with fidelity $99.95\%$ and $M$ ranging from $1$ to $10$. We show its adiabatic trajectory for $M=2$ in Fig. \ref{fig1} (a) and (b). The average adiabatic fidelity reaches $98\%$ during this adiabatic evolution process. $\vert W_M \psi_B\rangle$ can also be generated in $1.55\times2\pi\omega^{-1}$ with fidelity $99\%$, as depicted in Fig. 4 (c) of the main text.  We show its adiabatic trajectory for $M=3$ in Fig. \ref{fig1} (e) and (f). Finally, $\vert W'_2\rangle=\frac{1}{\sqrt{g_1^2+g_2^2}}(g_1\vert 10\rangle+g_2\vert01\rangle)$ can be generated in $1.86\times2\pi\omega^{-1}$ with fidelity $99\%$, as depicted in Fig. 4 (d) of the main text. We show its adiabatic trajectory for $g_1=2g_2$ in Fig. \ref{fig1} (g) and (h). Other adiabatic trajectories for generating different $\vert W_M\rangle$ with the same speed can be obtained by fixing $\sum g_{i}^2$ and tuning $g_i/g_j$.

\section{Effect of the environment and catch and release of the $W$ states}
\begin{figure}[h]
\centering
\includegraphics[scale=0.55]{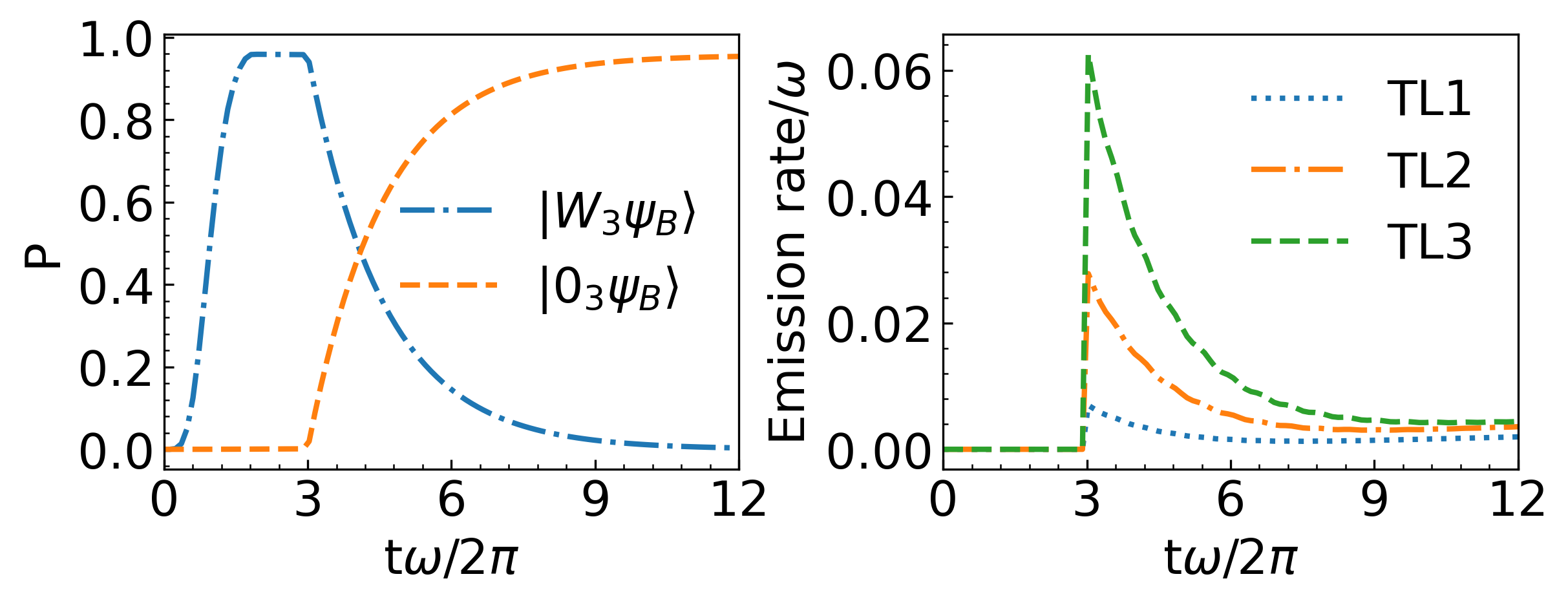}
\renewcommand\figurename{\textbf{FIG.}}
\caption[1]{Numerical simulation for the catch and release of the $W$ state $|W_3\rangle=1/\sqrt{14}(|100\rangle+2|010\rangle+3|001\rangle)$ by solving the master equation \eqref{lind}. (a) Population of different states during the process. (b) The photon emission rate of each mode into the corresponding TL.
}
\label{fig2s}
\end{figure}

We use the following Lindbladian master equation
\begin{eqnarray}\label{lind}
\dot{\rho}=&&-i[H_{pq},\rho]+\sum_{i=1}^M \frac{\kappa}{2}(2a_i\rho a_i^\dag- a_i^\dag a_i\rho-\rho a_i^\dag a_i)+\sum_{m=1}^2 \frac{\gamma_m}{2}(2\sigma_m\rho \sigma_m^\dag- \sigma_m^\dag \sigma_m\rho-\rho \sigma_m^\dag \sigma_m)\nonumber\\&&+\sum_{m=1}^2 \frac{\gamma_{m\phi}}{2}(\sigma_{mz}\rho \sigma_{mz} - \rho)
\end{eqnarray}
to study the effect of the environment and catch and release of the $W$ states, as shown in Fig. 5 of the main text. Here, $\kappa$ is the decay rate of the resonator, consisting of the intrinsic part $\kappa_{in}$ and variable coupling part $\kappa_c$ with respect to the transmission line (TL). $\gamma_m$  and $\gamma_{m\phi}$ are the energy relaxation rate and the pure dephasing rate of the $m$-th qubit, respectively. We choose $\kappa_{in}=10^{-4}\omega$, $\kappa_c=0.1\omega$, $\gamma_m=10^{-5}\omega$,  $\gamma_{m\phi}=2\times10^{-5}\omega$, where $\omega$ is the resonator frequency. Initially, the dissipation rate is very small, so that $|W_3\rangle=1/\sqrt{14}(|100\rangle+2|010\rangle+3|001\rangle)$ can be generated through adiabatic evolution along $\vert \psi_{2s+}\rangle$ in $1.86\times 2\pi\omega^{-1}$ by choosing $g_1:g_2:g_3=1:2:3$. $|W_3\rangle$ will be stored in the resonator till
$t=3\times 2\pi\omega^{-1}$. Then we turn on the variable coupling $\kappa_c$ to release the $W$ states into the TL. The emission rate into the $i$-th TL is $\kappa_c\langle a_i^\dag a_i\rangle$. This process is simulated by solving the master equation Eq. \eqref{lind} and depicted in Fig. \ref{fig2s}.

Normally, one should consider the following dressed master equation for the ultrastrong coupling regime \cite{dressed1,dressed2}
\begin{equation} 
\label{dressed}
\dot{\rho}(t) = -i[H,\rho(t)] + \sum_{j,k>j}(\sum_i^M \Gamma_{i\kappa}^{jk}+\Gamma_{\gamma_1}^{jk}+\Gamma_{\gamma_2}^{jk}) (D(\ket{j}\bra{k})\bm{\cdot} \rho(t)),
\end{equation}
\begin{figure}[h]
\centering
\includegraphics[scale=0.55]{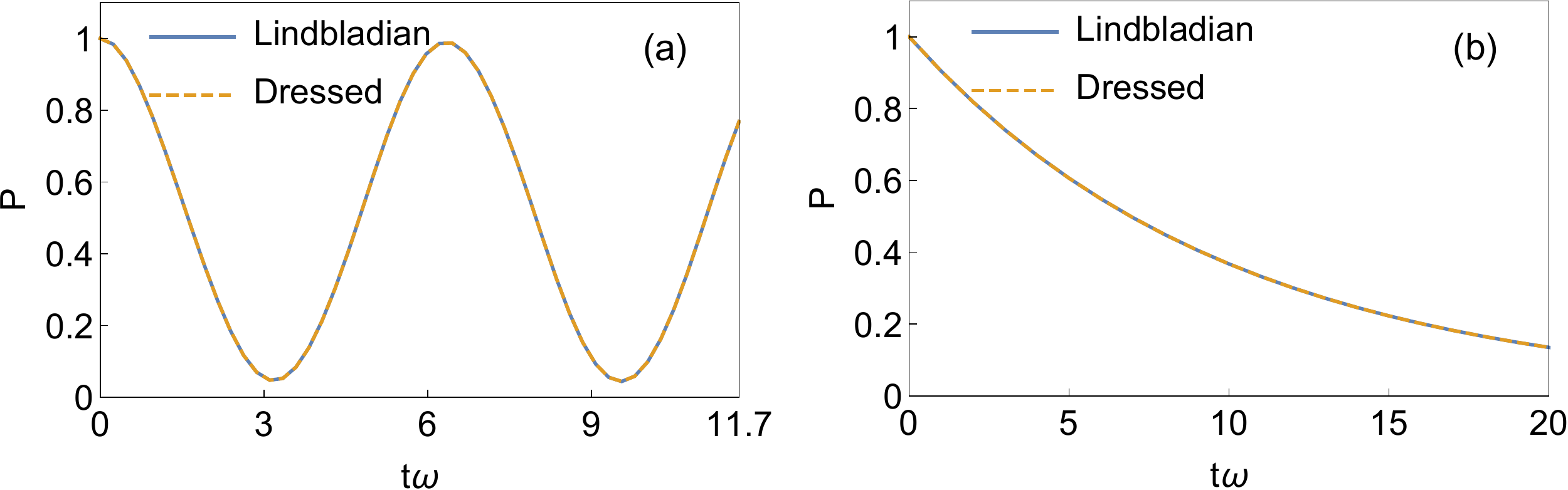}
\renewcommand\figurename{\textbf{FIG.}}
\caption{(a) Population of $\vert 0_3\uparrow\uparrow\rangle$ under the time-evolution generated by the two-qubit three-mode quantum Rabi model with $2\Delta_1=2\Delta_2=\omega$, $g_{11}=g_{12}=0.1\omega$, $g_{21}=g_{22}=0.2\omega$, $g_{31}=g_{32}=0.3\omega$, $\kappa=10^{-4}\omega$, $\gamma_1=\gamma_2=10^{-5}\omega$,  $\gamma_{1\phi}=\gamma_{2\phi}=0$. (b) Population of $\vert W_3\psi_B\rangle$ under the time-evolution generated by the two-qubit three-mode quantum Rabi model with $2\Delta_1=2\Delta_2=\omega_r=\omega$, $g_{11}=g_{12}=0.266\omega$, $g_{21}=g_{22}=0.532\omega$, $g_{31}=g_{32}=0.798\omega$, $\kappa=0.1001\omega$, $\gamma_1=\gamma_2=10^{-5}\omega$,  $\gamma_{1\phi}=\gamma_{2\phi}=0$. $\vert W_3\rangle=1/\sqrt{14}(\vert100\rangle+2\vert 010\rangle+3\vert001\rangle )$. } 
\label{fig2}
\end{figure}
where ${D(O)\bm{\cdot}\rho = 1/2(O \rho O^{\dagger} - O^{\dagger}O \rho - \rho O^{\dagger}O)}$, and $\{\ket{j} \}_{j=0,1,2..}$ are eigenvectors of the Hamiltonian $H$ with eigenenergy $\epsilon_{j}$. The decay rates read
\begin{eqnarray} 
\Gamma_{i\kappa}^{jk}&=& 
       \kappa \frac{\Delta_{kj}}{\omega}|\langle k|a_i+a_i^{\dag}|j\rangle|^{2},\\
\Gamma_{\gamma_m}^{jk} &=&\gamma_{m} \frac{\Delta_{kj}}{\omega_{qm}}|\langle k|\sigma_x|j\rangle|^{2},  \quad\  m=1,2,
\end{eqnarray}
where $\Delta_{kj} = \epsilon_{k} - \epsilon_{j}$. $\omega_{qm}$ is the frequency of the $m$-th qubit. However, the lindblad master equation is an appropriate approximation here.  The reason is as follows. Although the ultrastrong coupling regime is reached during the  adiabatic process to generate $\vert W_3\psi_B\rangle$, the dissipation rate is too small ($\kappa=10^{-4}\omega$, $\gamma_1=\gamma_2=10^{-5}\omega$) to significantly effect the system in a time of $1.86\times 2\pi\omega^{-1}$. $\kappa$ is increased to $0.1001\omega$ afterwards to emit the $W$ states. However, the qubits has become a singlet Bell state $\vert\psi_B\rangle=\frac{1}{\sqrt{2}}\vert\downarrow\uparrow-\uparrow\downarrow\rangle$, which is decoupled from the resonator. So the dissipation can be treated independently for the qubit and photon parts. We tested the validity of the Lindbladian master equation by comparing the dynamics obtained from Eqs. \eqref{lind} and \eqref{dressed}.
It can be seen from Fig. \ref{fig2} that the dressed master equation and Lindbladian master equation give almost the same results. We did not use the former to study the full dynamics because the Hamiltonian is changing with time, so that we have to diagonalize $H(t)$ and calculate many matrix elements  $\Gamma^{jk} $ every time, which costs much time.

\end{document}